\documentclass[%
 reprint,superscriptaddress,
 amsmath,amssymb,pra]{revtex4-1}
\usepackage{physics}
\usepackage{xcolor}
\usepackage{xtab,afterpage,longtable}
\usepackage{lipsum}
\usepackage{graphicx}
\usepackage{dcolumn}
\usepackage{booktabs} 
\usepackage{multirow}
\usepackage{color}
\newcommand*{\rom}[1]{\expandafter\@slowromancap\romannumeral #1@}
\usepackage{hyperref}

\begin{document}

\title{Robust AC vector sensing at zero magnetic field with pentacene}

\author{Boning Li}
    \thanks{These authors contributed equally.}
    \affiliation{Department of Physics, Massachusetts Institute of Technology, MA 02139, USA}
    \affiliation{Research Laboratory of Electronics, Massachusetts Institute of Technology, Cambridge, MA 02139, USA}
\author{Garrett Heller}
    \thanks{These authors contributed equally.}
    \affiliation{Research Laboratory of Electronics, Massachusetts Institute of Technology, Cambridge, MA 02139, USA}
    \affiliation{Department of Electrical Engineering and Computer Science, Massachusetts Institute of Technology, MA 02139, USA}
\author{Jungbae Yoon}
    \affiliation{Research Laboratory of Electronics, Massachusetts Institute of Technology, Cambridge, MA 02139, USA}
\author{Alexander Ungar}
    \affiliation{Research Laboratory of Electronics, Massachusetts Institute of Technology, Cambridge, MA 02139, USA}
    \affiliation{Department of Electrical Engineering and Computer Science, Massachusetts Institute of Technology, MA 02139, USA}
\author{Hao Tang}
    \affiliation{Department of material science and engineering, Massachusetts Institute of Technology, Cambridge, MA 02139, USA}
\author{Guoqing Wang}
\affiliation{International Center for Quantum Materials, School of Physics, Peking University, Beijing 100871, P.R. China}
\author{Patrick Hautle}
    \affiliation{Paul Scherrer Institute, 5232 Villigen, Switzerland}
\author{Yifan Quan}
    \affiliation{Department of Chemistry, University of Pennsylvania, Philadelphia, PA 19104, USA}
\author{Paola Cappellaro}%
    \email{pcappell@mit.edu}
    \affiliation{Department of Physics, Massachusetts Institute of Technology, MA 02139, USA}
    \affiliation{Research Laboratory of Electronics, Massachusetts Institute of Technology, Cambridge, MA 02139, USA}
    \affiliation{Department of Nuclear Science and Engineering, Massachusetts Institute of Technology, Cambridge, MA 02139, USA}
             
\begin{abstract}
Quantum sensors based on electronic spins have emerged as  powerful probes of microwave-frequency fields. Among other solid-state platforms,  spins in molecular crystals offer a range of advantages, from high spin density to  functionalization via chemical tunability. Here, we demonstrate microwave vector magnetometry using the photoexcited spin triplet of deuterated pentacene molecules, operating at zero external magnetic field and room temperature. We achieve  full three-dimensional microwave field reconstruction by detecting  the Rabi frequencies of anisotropic spin-triplet transitions associated with two crystallographic orientations of pentacene in naphthalene crystals. We further introduce a phase alternated protocol that extends the rotating-frame coherence time by an order of magnitude and enables sensitivities of $1~\mu\mathrm{T}/\sqrt{\mathrm{Hz}}$ with sub-micrometer spatial resolution. These results establish pentacene-based molecular spins as a practical and high-performance platform for microwave quantum sensing, and the control techniques are broadly applicable to other molecular and solid-state spin systems.
\end{abstract}

\maketitle

\section{Introduction}
Quantum sensors based on optically-detected solid-state defect spin qubits have demonstrated remarkable sensitivity to electromagnetic fields~\cite{Taylor2008, Maze2008, Pham2011, Neumann2013, Wang2023}, enabling nanoscale imaging and precision metrology for spin-wave imaging~\cite{lee2020nanoscale,van2015nanometre}, micro-electronic device characterization~\cite{lillie2019imaging, nowodzinski2015nitrogen}, and bio-sensing \cite{tan2022emerging, miller2020spin}. Despite these advances, defect-based quantum sensors face intrinsic limitations. Their scalability is constrained by the difficulty of creating high-density, uniform defect ensembles, as the accompanying increase in surrounding spin impurities leads to decoherence~\cite{Zhou2020, Ghassemizadeh2024,bauch2020decoherence,park2022decoherence,Mittiga2018,wang2022sensing}. Furthermore, at low external magnetic fields, inhomogeneous charge and strain environments mix the spin energy levels, degrading both the coherence and the sensing accuracy~\cite{zhu2014observation,matsuzaki2016optically,matsuzaki2015improving}.

Molecular quantum systems \cite{Kohler1993, Wrachtrup1993} with optically addressable spin degrees of freedom have recently emerged as a compelling alternative route for quantum sensing~\cite{Singh2025, Mena2024,singh2025high,yamauchi2024modulation}. These molecular spins can exhibit long coherence times and combine the advantages of chemically programmable spin sites with high density doping in many host crystals~\cite{Zadrozny2017, Bayliss2020}. 
Among these systems, \textit{pentacene} stands out for its long-lived photoexcited triplet states and efficient optical spin polarization at room temperature~\cite{Singh2025, Mena2024, Kohler1993,Wrachtrup1993}. Recent studies incorporate pentacene into thin films~\cite{Moro2022, Lin1997}, patterned directly on chip~\cite{Yunus2022}, or grown as bulk crystals, demonstrating versatility for device integration and scalable sensing architectures~\cite{zhang2024ultra,wu2022enhanced}.

Here, leveraging an ensemble of photoexcited triplet states of pentacene molecules embedded in a naphthalene single crystal, we develop a Rabi-based protocol that exploits the intrinsic multi-orientational structure and anisotropic spin transitions for vector AC magnetic-field sensing at microwave frequencies. To further enhance its sensitivity, we introduce a phase-modulated control scheme~\cite{Solomon1959,Aiello2013,hirose2012continuous} that effectively decouples the qubit from field fluctuations and inhomogeneities that arise from the environment and the driving itself. The magnetic-field sensitivities reaches $1~\mu\mathrm{T}/\sqrt{\mathrm{Hz}}$  with sub-micrometer spatial resolution.

\begin{figure*}[htbp]
\includegraphics[width=1\textwidth]{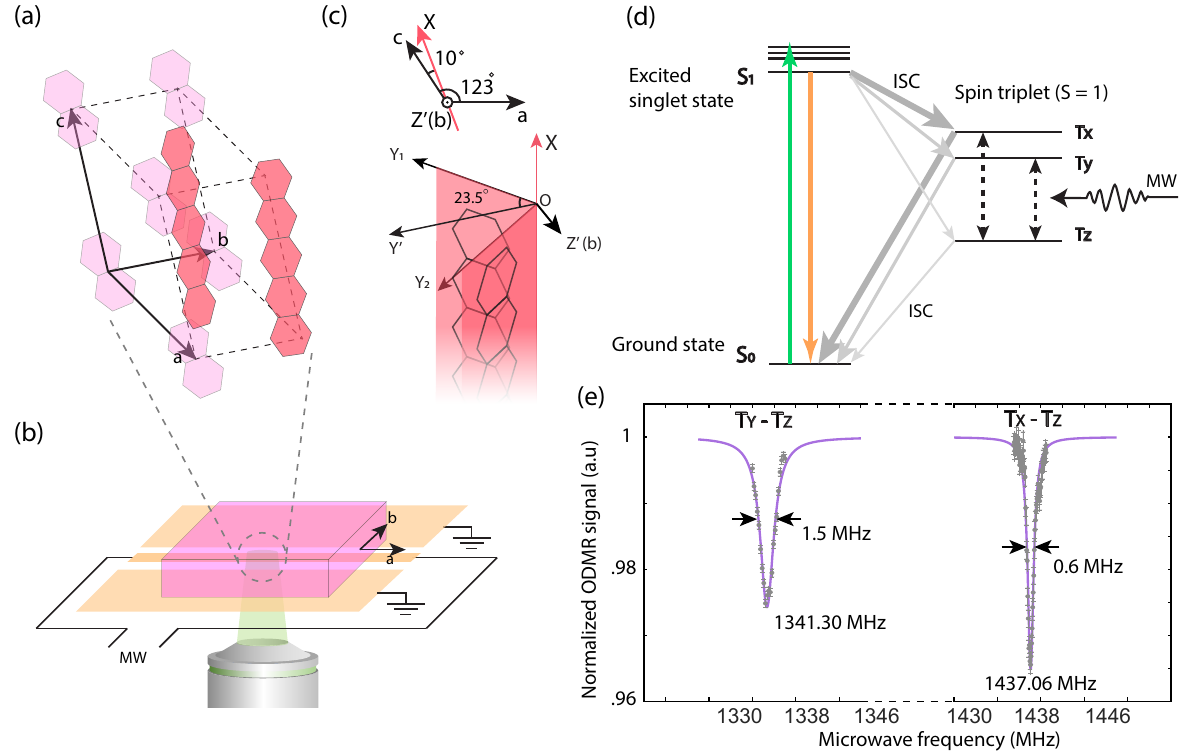}
\caption{\textbf{Experimental setup and pentacene properties.} 
(a) Crystal structure of pentacene-doped naphthalene, where $a$, $b$, and $c$ denote the crystallographic axes. The cleaved surface corresponds to the $ab$-plane. 
(b) Pentacene molecules substitute into two inequivalent lattice sites of the naphthalene host, giving rise to two crystallographically distinct molecular orientations. Both orientations share a common molecular long axis (defined as the $X$-axis), which lies in the $ac$-plane and is tilted by approximately $10^{\circ}$ from the $c$-axis. The local molecular frames are denoted $X OY_1$ and $X O Y_2$, where $Y_1$ and $Y_2$ lie along the short molecular axis within the molecular plane. The Hamiltonian in Eq.~\eqref{eq:hamiltonian_mole} is expressed in this molecular frame. The $Y_1-$ and $Y_2-$  axes are symmetrically split about the $ac$-plane by $47^{\circ}$. For convenience, we define an orthonormal sensing frame $X Y' Z'$, where $X$ coincides with the molecular long axis, $Z'$ is aligned with the crystal $b$-axis, and $Y'$ bisects the angle between $y_1$ and $y_2$.  
(c) Schematic of the experimental setup showing the crystal cleavage plane ($ab$-plane). The $a$- and $b$-axes can be identified independently~\cite{Quan2023}. Microwave excitation is delivered using a coplanar-waveguide stripline, and optical excitation (532~nm, 0.1~mW) and readout are performed with a confocal microscope.
(d) Energy-level diagram of pentacene and relevant inter-level transitions. Green and orange arrows denote optical excitation and fluorescence, respectively, while gray arrows represent the non-radiative intersystem crossing (ISC) processes. The ISC coupling to the $T_X$ sublevel is stronger than to $T_Y$ or $T_Z$. Microwave fields can drive transitions between the triplet spin-states. 
(e) Optically detected magnetic resonance (ODMR) spectra of the $T_X \leftrightarrow T_Z$ and $T_Y \leftrightarrow T_Z$ transitions measured at zero magnetic field. The linewidths are obtained from Lorentzian fitting.
\label{fig:1}
}
\end{figure*}

\section{Room-temperature vector AC sensing}
\subsection{{Pentacene sample and spin properties}}

We used pentacene-d$_{14}$-doped naphthalene single crystals  with a pentacene concentration of about  $4\times10^{-5}$ mol/mol~\cite{Quan2023}. The sample was extensively purified to achieve high quality single crystals and minimize impurities (see supplementary materials \cite{supp} for detailed sample preparation and characterization).
The pentacene molecules align along two crystallographically distinct orientations from substitution of the naphthalene host at inequivalent lattice sites [Fig.~\ref{fig:1}(a)]. These orientations are fixed relative to the crystal axes, thereby establishing a well-defined geometric relation between the molecular frames and the laboratory sensing frame [Fig.~\ref{fig:1}(b)]. 

The naphthalene lattice vectors $\vec{a}$ and $\vec{b}$ define the cleavage ($ab$) plane, which can be experimentally identified through birefringence or angle-resolved optical measurements~\cite{Quan2023}. In our confocal microscope setup, the laboratory sensing frame is aligned with respect to this crystallographic axis system, as illustrated in Fig.~\ref{fig:1}(c).

At zero external magnetic field, the two oriented of pentacene molecules are energetically degenerate.
An electronic spin-triplet metastable state~\cite{Singh2025,Mena2024} presents in a pentacene molecule, as shown in the simplified energy-level structure [Fig.~\ref{fig:1}(d)]. Under optical excitation, population can be pumped from the singlet ground state ($S_0$) to the excited singlet state ($S_1$), from which it can either decay radiatively back to $S_0$ or undergo spin-dependent intersystem crossing (ISC) into the triplet manifold ($T_X$, $T_Y$, $T_Z$). 
The spin-dependent ISC process preferentially populates the $T_X$ state, enabling optical spin polarization. To prepare an initial triplet polarized on other sublevels, a fast ($\sim$100~ns) flip microwave pulse is employed immediately after laser pumping.

The triplet state decay to the ground state $S_0$ also via spin-dependent ISC, with decay rates $\Gamma_\mu$ ($\mu=X,Y,Z$) that are unequal among the triplet sublevels. Therefore, the population of the $S_0$ state after a delay time $t_d$ encodes information about the previous excited spin state. 
The resulting photoluminescence (PL), which is proportional to the population in $S_0$, thus provides spin-dependent contrast and enables optical readout of  the triplet manifold. 

We define the orthogonal molecular frame such that the $X$- and $Y$-axes align with the long and short molecular axes of an individual pentacene molecule, respectively (frames $XOY_1$ and $XOY_2$ in Fig.~\ref{fig:1}(b), noting that the two orientations share the same $X$–axis). In this molecular frame, the triplet energy levels are described by the Hamiltonian
\begin{equation}
    \mathcal{H} = D S_Z^2 - E (S_X^2 - S_Y^2),
    \label{eq:hamiltonian_mole}
\end{equation}
with zero-field splitting parameters $D = 1389~(2\pi)\mathrm{MHz}$ and $E = -48~(2\pi)\mathrm{MHz}$~\cite{Quan2021}.  
Transitions between the triplet eigenstates can be driven by a microwave field $\vec{B}_{\mathrm{ac}}(t)\!\cdot\!\vec{S}$ through the nonzero matrix elements
\begin{equation}
    \langle T_X | S_Z | T_Y \rangle
    = \langle T_Z | S_Y | T_X \rangle
    = \langle T_Y | S_X | T_Z \rangle
    = 1 .
    \label{eq:matrix_op}
\end{equation}
Therefore, any pair of triplet sublevels can serve as a qubit.  
To correct for imperfect state preparation and optical collection, we normalize the raw PL signal by a reference PL measured at the same delay $t_d$ after initializing the system (e.g.\ $\ket{T_\mu}$). Assuming we operate in a qubit manifold with populations $|\alpha|^2+|\beta|^2=1$ for the two states $\mu,\nu$, the normalized signal is thus 
\begin{equation}
  \mathrm{Sig} 
    = 1-|\beta|^2
      \frac{e^{-(\Gamma_\nu t_d)^{\iota_{\nu}}} - e^{-(\Gamma_\mu t_d)^{\iota_{\mu}}}}
           {1 - e^{-(\Gamma_\mu t_d)^{\iota_{\mu}}}}
           = 1- |\beta|^2 C(t_d),
    \label{eq:contrast}
\end{equation}
where the stretch exponents $\iota_{\mu,\nu}$ capture the multichannel relaxation pathways. This defines the readout contrast $C(t_d)$. Experimentally, $t_d = 30 ~\mu\mathrm{s}$ and $47 ~\mu\mathrm{s}$ maximize the contrast for the $T_X \leftrightarrow T_Z$ and $T_Y \leftrightarrow T_Z$ qubits, respectively.  Figure~\ref{fig:1}(e) shows the optically detected magnetic resonance (ODMR) of the two transitions at zero magnetic field. More details are presented in Supplementary materials.
\begin{figure}[htbp]
\includegraphics[width=0.46\textwidth]{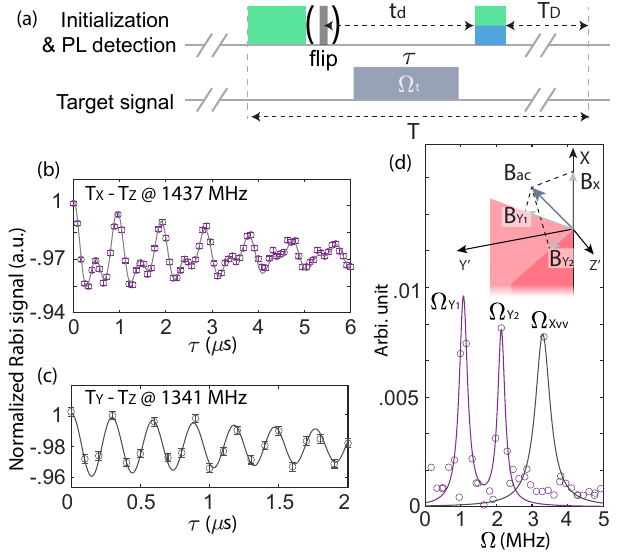}
\caption{\textbf{Vector AC sensing using two pentacene orientations.} 
(a) Experimental pulse sequence for ODMR and Rabi sensing. The green and blue blocks denote optical initialization and photon collection, respectively. An optional, short ($\sim100$~ns) flip pulse maps the initial $T_X$ state to $T_Y$ when we target  the $T_Y \leftrightarrow T_Z$ transition. After applying the continuous target driving $\Omega_t$, optical readout is performed following a delay time $t_d$, during which the triplet sublevels relax at different rates, generating fluorescence contrast between spin states. A post-detection delay time $T_D$ is needed after each PL measurement to reset the population, yielding  the total sequence time $T\approx 500\mu$s. (b–d) Optically detected Rabi oscillations of the $T_X \leftrightarrow T_Z$ (b) and $T_Y \leftrightarrow T_Z$ (c) transitions, and their corresponding Fourier spectra (d). Purple and gray curves show the $T_X \leftrightarrow T_Z$ and $T_Y \leftrightarrow T_Z$ transitions, respectively. The Rabi frequencies are proportional to the projection of the applied AC magnetic field onto the relevant molecular axis (inset), e.g., $\Omega_{Y_1} = \gamma_e B_{Y_1}$. Error bars reflect photon shot noise and are obtained by propagating the standard deviation of the measured photon counts through  signal normalization.
}
\label{fig:2}
\end{figure}

\subsection{Principle and experimental demonstration}
According to Eq.~\eqref{eq:matrix_op}, since each transition between pairs of triplet states couples to a specific component of the microwave magnetic field, the corresponding Rabi frequency encodes directional information of an external AC field. Thus, it would be possible to perform vector AC magnetometry at the level of a single pentacene molecule.  

A more compact protocol can be obtained exploiting the different pentacene orientations in the crystal, similar to prior schemes with NVs~\cite{wang2015high, Schloss2018}.
As mentioned above, the two pentacene orientations share a common molecular $X$-axis but have distinct $Y$-axes ($Y_1$ and $Y_2$). Therefore, when an AC microwave field is applied, the $T_Y \leftrightarrow T_Z$ transition in both orientations is driven equally by the $X$-component of the field, whereas the $T_X \leftrightarrow T_Z$ transition probes the orientation-dependent $y$-components ($Y_1$ and $Y_2$), as illustrated in the inset of Fig.~\ref{fig:2}(d).
Consequently, Rabi measurement yields two Rabi frequencies ($\Omega_{Y_1}$, $\Omega_{Y_2}$) from $T_X \leftrightarrow T_Z$ transitions in two orientations [Fig.~\ref{fig:2}(b,d)], and one ($\Omega_X$) from the $T_Y \leftrightarrow T_Z$ transition [Fig.~\ref{fig:2}(c,d)]. This three-axis projection scheme enables full reconstruction of the three-dimensional AC magnetic-field vector. For simplicity, we define the orthogonal laboratory sensing frame $XY'Z'$, where the $Y'$ axis bisects the molecular $Y_1$ and $Y_2$ axes [Fig.~\ref{fig:1}(b) and Fig.~\ref{fig:2}(d), inset] and $Z'$-axis aligns with the $b$ axis of naphthalene lattice. The AC magnetic-field components in this frame can be reconstructed as
\begin{equation}
    \begin{aligned}
        \gamma_eB_{X}  &= \Omega_{X}, \\
         \gamma_eB_{Y'} &= \frac{\Omega_{Y_1} +\Omega_{Y_2}}{2\cos{\phi}}, \\
         \gamma_eB_{Z'} &= \frac{\Omega_{Y_1} -\Omega_{Y_2}}{2\sin{\phi}},
    \end{aligned}
    \label{eq:vector}
\end{equation}
where  $\gamma_e$ is the electron spin gyromagnetic ratio of pentacene, equal to that of a free electron.

The Rabi oscillation appears as a damped oscillation signal with a characteristic decay time $T_{2\rho}$. Considering both photon shot noise and the finite spin coherence time,  the sensitivity $\eta$ of the Rabi protocol is given by~\cite{degen2017quantum,barry2020sensitivity}
\begin{equation}
    \eta = 
    \frac{\sigma_S}{C(t_d)}  
    \frac{ e^{t/T_{2\rho}} }{ \gamma_e  t}\sqrt{T},
    \label{eq:sens_ori}
\end{equation}
where $t$ is the interrogation time and $C(t_d)$  the optical contrast from equation \ref{eq:contrast}, with waiting time $t_d$. $T = 500~\mu\mathrm{s}$ denotes the total experimental cycle time, which includes the sequence duration and a post-delay time $T_d$ that resets the population before each repetition of the sequence. [Fig.~\ref{fig:2}~(a)].  $\sigma_S$ represents the relative standard deviation of the normalized signal and is typically governed by photon shot noise.  The optimal interrogation time is set by the Rabi decay time,  $t \approx T_{2\rho} \approx 3~\mu\mathrm{s}$ for an AC magnetic field with amplitude $|B_{\mathrm{ac}}/\gamma_e| \sim 0.5~(2\pi)\mathrm{MHz}$. This corresponds to a sensitivity of  $\eta \approx 3.6~\mu\mathrm{T}/\sqrt{\mathrm{Hz}}$ for all three molecular axes. Notably, in the relevant regime where $T_{2\rho} \ll T$, the optical readout sensitivity scales as $\eta \propto 1/T_{2\rho}$. 

While we demonstrated the feasibility of optically detected molecular-spin–based vector magnetometry, the performance, especially the sensitivity of the simple Rabi protocol, remains substantially below that of other solid-state spin sensors. In the following, we analyze the sources of noise that limit the coherence time in order to devise a control strategy to improve the sensor performance.
\begin{figure}[htbp]
\includegraphics[width=0.46 \textwidth]{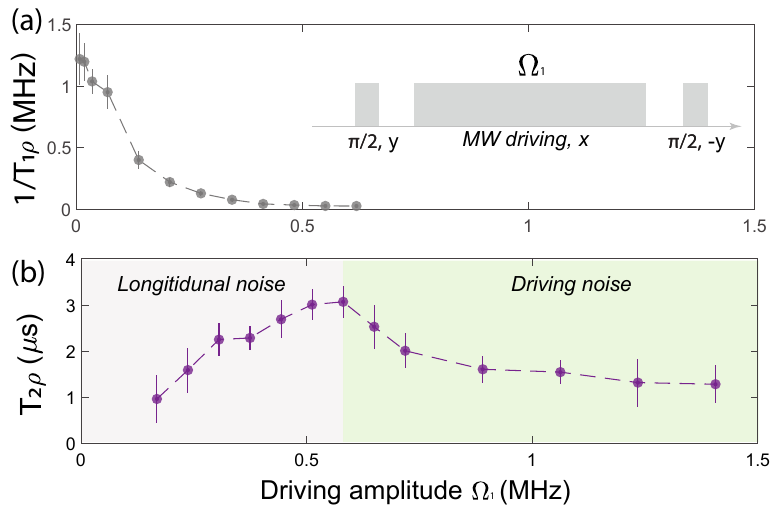}
\caption{\textbf{Pentacene coherence under continuous driving.} (a) Microwave pulse sequence for spin locking and measured longitudinal relaxation times $T_{1\rho}$ at varying driving amplitudes. Each point results from  fitting an exponential decay on the spin-locking signal for varying  MW driving duration. The decay rate $1/T_{1\rho}$ reflects the longitudinal noise spectrum $S_Z(\nu)$ evaluated at the driving strength. 
(b) Rabi relaxation time $T_{2\rho}$ (sequence in Fig.~\ref{fig:2}.a) as a function of the driving amplitude. The decrease in $T_{2\rho}$ at weak drive arises from magnetic-field fluctuations and hyperfine-induced spectral broadening, while the reduction at strong drive is dominated by microwave-amplitude noise. Error bars represent the 95\% confidence intervals of the  damped Rabi oscillation fit.
}
\label{fig:3}
\end{figure}
\section{Robust modulated AC sensing}
\begin{figure*}[htbp]
\includegraphics[width=1\textwidth]{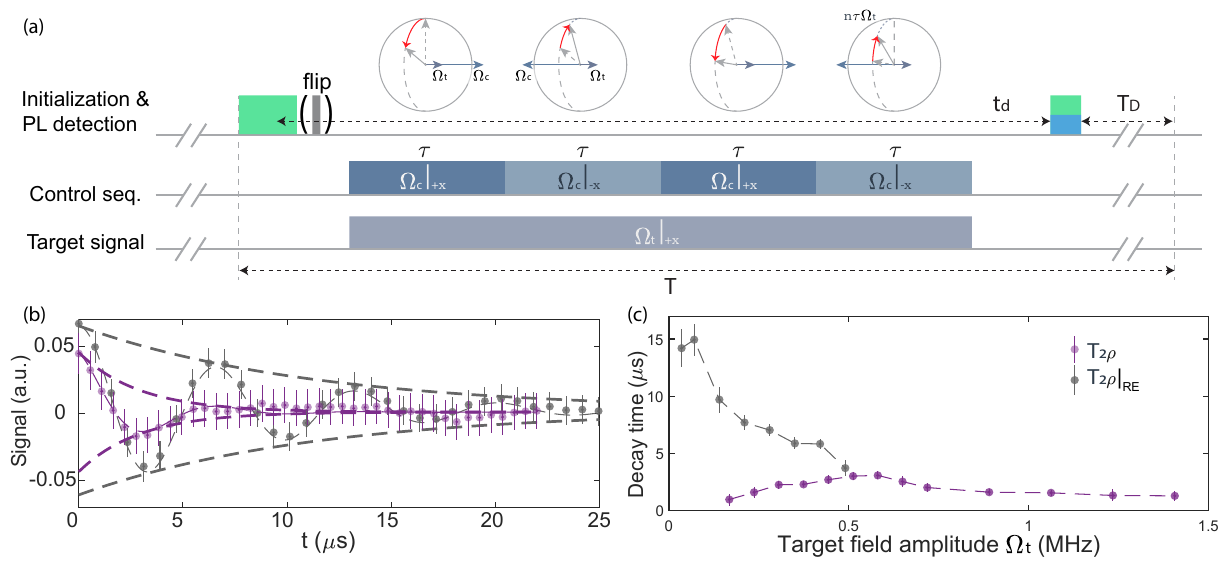}
\caption{\textbf{Rotary-echo sensing protocol.} (a) Experimental sequence for sensing. The  RE control sequence consists strong microwave driving of amplitude $\Omega_c$ with $\pi$ phase flips after each multiple of $\tau$. The target signal to be sensed is continuously applied $\Omega_t$. The flip pulse is still for $T_Y\leftrightarrow T_Z$ transition only. (b) Rabi oscillations without (purple) and with (gray)  the control sequence on the
$T_X \leftrightarrow T_Z$ transition performed at a target Rabi frequency of $0.17  ~(2\pi) \mathrm{MHz}$. Here the normalized signals are centered around zero for clear comparison of the coherence and contrast between the two scenarios. (c) Decay time, $T_{2\rho}$, versus target field amplitude, $\Omega_t$, in the case of no-control and RE
(with $\Omega_c = 2.625 ~(2\pi)\mathrm{MHz}, \tau = 390 ~\mathrm{ns}$). Note that in the case of weak driving, the RE control vastly lengthens the coherence time--about an order of magnitude longer than the simple Rabi driving.
}
\label{fig:4}
\end{figure*}
\subsection{Decoherence sources in driven pentacene spin systems}
The dominant sources of decoherence arise from the nuclear spin bath, from slowly varying spatial inhomogeneities of the static field, and from variations in the microwave amplitude due to electronic noise or spatial inhomogeneity of the drive. 
We can thus model the noise as a stochastic magnetic field that is added to the driving  Hamiltonian, 
\begin{equation}
H(t)
= \Omega_{1}\!\left[1+\xi_{\Omega_1}(t)\right]\cos(\omega_0 t)\sigma_X
+ \xi_Z\sigma_Z
+ \xi_X\sigma_X,
\label{eq:H_noise_disc}
\end{equation}
where $\xi_j$ denotes a stochastic magnetic field with power spectral density (PSD) $S_j(\nu)
= \int_{-\infty}^{\infty}\!\mathrm{d}t\,e^{i\nu t}\langle\xi_j(t)\xi_j(0)\rangle,
\  j\in\{z,x,\Omega_1\}$. 
Here $\xi_Z(t)$ and $\xi_X(t)$ are longitudinal and transverse stochastic magnetic fields, respectively, and $\xi_{\Omega_1}(t)$ represents relative fluctuations in the drive amplitude.

When the goal is to sense weak AC fields, $\xi_{\Omega_1}$ can be neglected and longitudinal inhomogeneous broadening limits the coherence time. In deuterated pentacene, the observed $\sim 0.6 ~(2\pi)\mathrm{MHz}$\, linewidth is dominated by hyperfine coupling to the deuterons in the pentacene molecule~\cite{Singh2025,Quan2021}. This coupling introduces a distribution of quasi-static detunings $\delta\omega$ that yield an effective Rabi frequency
\(
\Omega_R(\delta\omega)=\sqrt{\Omega_1^2+\delta\omega^2}.
\)
At low microwave amplitudes ($\Omega_1 \lesssim \delta\omega$), the finite linewidth introduces a bias in the Rabi oscillation frequency, faster dephasing, and reduces the contrast~\cite{supp}, thereby degrading the sensing accuracy~\cite{billaud2025electron,Kitamura2025}. 
A simple strategy to mitigate the broadening effects would be to add a bias driving field to the target AC, in order to operate at larger driving strengths. When
\(
\langle \Omega_R \rangle \approx \Omega_1 +{\langle \delta\omega^2 \rangle}/{2\Omega_1}
\), detuning-induced decoherence and bias is strongly suppressed.

In this strong-driving limit, relaxation can be quantified by the generalized Bloch-equation (GBE) framework~\cite{geva1995relaxation,Wang2020}. Moving to the rotating frame at resonant frequency $\omega_0$, the longitudinal and transverse (Rabi) relaxation rates are given by 
\begin{equation}
\begin{aligned}
    \frac{1}{T_{1\rho}} &= S_Z(\Omega_1) + \frac{1}{2}S_X(\omega_0),\\[4pt]
    \frac{1}{T_{2\rho}} &= \frac{1}{2T_{1\rho}}
    + \frac{1}{2}S_X(\omega_0)
    + \frac{1}{4} S_{\Omega_1}(0).
\end{aligned}
\label{eq:T1rho_T2rho_disc}
\end{equation}

The relaxation time $T_{1\rho}$ depends marginally on the transverse noise, as it is only affected by spectral density components, $S_X(\omega_0)$, around the high frequency $\omega_0$. Thus, the longitudinal relaxation is governed by $S_Z(\Omega)$, longitudinal noise spectral components at frequency around the dressed-state splitting $\Omega_1$. We determine $S_Z(\nu)$ by measuring $T_{1\rho}$ with a spin-locking experiment \cite{Abragam1961,li2025exploring} see insets of Fig.~\ref{fig:3}~(a). The results show that the spectrum is zero-frequency centered, consistent with slow fluctuations from the nuclear-spin environment. Combined with the hyperfine-induced inhomogeneous broadening analyzed above, this explains both the rapid decoherence at weak drive and its suppression as the drive amplitude increases.

However, as we increase the driving amplitude $\Omega_1$  the effects of its inhomogeneity and fluctuation on the Rabi coherence increases, see Fig.~\ref{fig:3}~(b), since $S_{\Omega_1} \propto \Omega_1$~\cite{Wang2020}. For slow varying amplitude fluctuations (with correlation time $\gg$ interrogation time), the decoherence exhibits a Gaussian decay envelope with a rate proportional to $\Omega_1$ (detailed derivation in the supplementary materials~\cite{supp}.)

Taken together, our experimental results and noise model reveal a fundamental limitation to the bandwidth and sensitivity of Rabi-based sensing. At low drive amplitudes, the sensing performance degrades due to inhomogeneous detuning broadening and quasi-static magnetic noise. At high drive amplitudes, the coherence becomes limited by microwave inhomogeneity, consistent with the observed variation of $T_{2\rho}$ across different driving conditions.

\subsection{Phase alternation for enhanced Rabi sensing}
To extend the driven-state coherence and enable robust Rabi-type sensing, we adopt a rotary-echo (RE) scheme~\cite{Solomon1959} to protect the coherence of the continuous target field drive. The protocol applies a strong resonant drive whose phase is periodically inverted at intervals $\tau$ [Fig.~\ref{fig:4}(a)], thus canceling drive imperfections. Similar  coherence protection is achieved with more complex piece-wise constant~\cite{Levitt1981, Levitt1982, Shaka1983, Shaka1988} or continuous phase modulation, such as  by concatenated continuous dynamical decoupling (CCDD)~\cite{Cai2012, Wang2020}. Such techniques have also been applied to coherent sensing in the radio-frequency band~\cite{Wang2021, wang2022sensing,Aiello2013, hirose2012continuous}. Recently, CCDD-based protocols~\cite{Stark2017,Salhov2024,Kitamura2025} and frequency mixing~\cite{wang2022sensing} were proposed to improve microwave sensing, but their performance was still limited by a reduction of the target field amplitude or by drive-induced inhomogeneities. Here, we show that our rotary-echo approach effectively refocus drive-induced errors while preserving the desired decoupling from external magnetic fields.

In our scheme, we implement a periodic train of rotary echoes with alternating phases, where the protecting drive is always oriented with the target field,  ensuring that the evolution generated by the control is properly time-reversed. After an even number of inversions, the qubit recovers a clean Rabi oscillation governed solely by the transverse target field. The resulting Rabi frequencies can then be directly used for the microwave–field vector reconstruction discussed in the previous section.

Experimentally, we observe a substantial extension of the driven-state coherence time under the RE control. 
As shown in Fig.~\ref{fig:4}(b), under a weak target field ($\Omega_{\mathrm{t}} \sim 0.2~(2\pi)\mathrm{MHz}$) between the $T_X \leftrightarrow T_Z$ transition, the simple Rabi protocol produces barely visible oscillations, whereas the RE sequence with strong driving ($\Omega_\mathrm{c} = 2.625 ~(2\pi)\mathrm{MHz}$, $\tau = 390~\mathrm{ns}$) restores clear, long-lived coherence. The coherence time as a function of the target-field amplitude is shown in Fig.~\ref{fig:4}(c), demonstrating that RE extends the driven-state coherence by nearly an order of magnitude compared to the standard Rabi protocol.

A similar improvement is also observed for the $T_Y \leftrightarrow T_Z$ transition when the RE control is applied (\cite{supp}). 
The same vector-field reconstruction procedure described in Eq.~\eqref{eq:vector} can be implemented under the RE control. The sensitivity for the vector field with extended $T_{2\rho}$ is $\left(\eta_X,\eta_{Y'},\eta_{Z'}\right) = \left(0.8, 0.5,1.1\right)~\mu T/\sqrt{\text{Hz}}$, comparable to solid-state spin sensors operated under finite static magnetic fields~\cite{wang2015high}. 

\subsection{Coherence protection mechanisms}
To elucidate the protection offered by the RE scheme and identify optimal driving parameters, we analyze the driven-spin dynamics still using the GBE framework.

The Hamiltonian in Eq.~\eqref{eq:H_noise_disc} is modified by substituting the term proportional to $\Omega_1$ with
\begin{equation}
H_{\mu w}
= \frac{1}{2}\!\left[
\Omega_{\mathrm{t}}
+ \Omega_{\mathrm{c}}\, s(t)\big(1+\xi_{\Omega_c}(t)\big)
\right]\sigma_X,
\label{eq:H_RCPMG_disc}
\end{equation}
where we neglected noise from the target field. Here $\xi_{\Omega_c}(t)$ denote relative fluctuations of the control amplitudes, and $s(t)=\pm 1$ implements the periodic inversions with period $2\tau$. 

The RE control modifies the dressed-state dynamics in two key ways.  First, dephasing arising from amplitude fluctuations of the control field are modulated by the periodic inversion function $s(t)$ and efficiently suppressed. The modulation generates a filter function that modifies the response to the noise spectrum $S_{\Omega_c}(\nu)$ from zero frequency to the odd harmonics of the modulation frequency $\pi/\tau$~\cite{alvarez2011measuring,bylander2011noise}. The resulting dephasing rate contributed by the control-field amplitude noise is centered around these higher harmonics,
\begin{equation}
    \Gamma_{\Omega_{c}}
    \approx 
    \sum_{m=-\infty}^{\infty}
    \frac{1}{(2m+1)^2}
    S_{\Omega_c}\!\left[\frac{(2m+1)\pi}{\tau}\right],
\end{equation}
and thus effectively suppressed.

Second, the phase modulation modifies the longitudinal-noise decoupling and care should be taken to ensure that it is still effective. Indeed, the periodic phase inversions generate Floquet sidebands in the dressed-state splitting~\cite{Aiello2013, hirose2012continuous,ivanov2021floquet}, modifying the frequency components of the noise that satisfy the dressed-state resonance condition. In the strong-control limit ($\Omega_{\mathrm{c}}\gg\Omega_{\mathrm{t}}$),  the longitudinal relaxation due to the $\xi_Z$ is:
\begin{equation}
\left.\frac{1}{T_{1\rho}}\right|_{\mathrm{R\text{-}SE}}
\simeq 
\sum_{m=-\infty}^{\infty}
\frac{1}{(2m+1)^2}
\,S_Z\!\left[
\Omega_{\mathrm{c}} - \frac{(2m+1)\pi}{\tau}
\right].
\label{eq:T1rho_RCPMG_Floquet}
\end{equation}
By choosing sufficiently large $\Omega_{\mathrm{c}}$ and appropriate inversion spacing $\tau$, the dominant Floquet sideband frequencies are shifted far from zero. Experimentally [Fig.~\ref{fig:4}(b,c)], we set $\Omega_{\mathrm{c}}\tau = 2\pi$, which places the dominant coupling near $\Omega_{\mathrm{c}}/2,$ well separated from the zero-frequency peak of $S_Z(\nu)$. This effectively decouples the dressed states from low-frequency bath fluctuations.

The  RE protocol provides an effective biasing scheme to enhance Rabi-based sensing, enabling accurate weak-field detection and enhancing the vector sensitivities demonstrated in this work.

\section{Discussion and Outlook}
In this work, we propose and experimentally demonstrate a protocol that employs photoexcited pentacene triplet spins in a naphthalene crystal as a vector AC magnetometer operating at room temperature and zero static field. The three-dimensional components of an AC magnetic field are reconstructed from the Rabi frequencies of selected transitions from the two crystallographic pentacene orientations in naphthalene crystal. The insights gained by experimentally evaluating and theoretically modeling the noise affecting the sensor lead us to introduce the  rotary-echo quantum sensing protocol, which significantly improves sensitivity by protecting the spin coherence.
The proof-of-concept implementation in this work achieves sensitivities comparable to established solid-state defect sensors, on the order of $\sim 1~\mu\mathrm{T}/\sqrt{\mathrm{Hz}}.$ 

Experimental evidence indicates an inherent bandwidth–sensitivity trade-off for standard Rabi sensing. The longitudinal noise limits the sensitivity in low field regime, while the inhomogeneity of the strong drive introduces additional dephasing. The  RE sequence applies strong driving with periodic phase inversions which refocuses the inhomogeneous drive amplitudes, thus providing a perfect biasing of the weak field to the strong driving regime and effectively decouples to the noise field. The coherence enhancement is particularly important for pentacene-based protocols, which necessitates a substantial overhead time to build spin-state contrast, causing the overall sensitivity to scale linearly with the available coherence. Recent advances in rotating-frame coherence protection for NV centers~
\cite{Stark2017,Salhov2024,Kitamura2025} may inspire future studies on their applicability to pentacene. 

For general photoexcited molecular spin systems, the long overhead time is often a disadvantage with respect to ground-state solid-state defect sensors. However, as we demonstrate, molecular ensembles can achieve comparable sensitivities, benefiting from their significantly higher spin densities while preserving long coherence times. Throughout our experiments,  $\sim 60\%$ of the spins in the detection volume were optically excited~\cite{supp}. Practically, this initialization efficiency could have 1.25$\times$ increase when using a $552\,\mathrm{nm}$ pump laser, corresponding to the room temperature absorption maximum of pentacene~\cite{Quan2023}, which could further improves the sensitivity. 

An important advantage of using pentacene molecular spins to perform AC magnetometry is the possibility to work at zero external magnetic field. In this regime, the accuracy and sensitivity of typical solid-state defect sensors degrade due to local strain, charge and external defects that induce state mixing and decoherence~\cite{Vetter2022, Zheng2019, Udvarhelyi2018, Ungar2025,Mittiga2018,zhu2014observation,matsuzaki2016optically,matsuzaki2015improving,saijo2018ac}. In contrast, the vast majority of impurities and lattice defects can be removed through extensive purification of organic crystals.
Additionally, the  triplet level structure of pentacene is intrinsically robust to small background magnetic fields, whose influence is quadratically suppressed (e.g., $<20~(2\pi)\,\mathrm{kHz}$ resonance shifts under Earth magnetic field). This  insensitivity makes pentacene a practically convenient platform in unshielded or imperfectly controlled environments.

Looking forward, molecular spin systems offer unique opportunities for quantum applications, benefiting  from chemical engineering, including nuclear-spin~\cite{Atzrodt2018,ryan2025spin} or ligand modification~\cite{Graham2017,mirzoyan2021deconvolving,wedge2012chemical}, as well as programmable spatial placement of the spin hosts~\cite{Lavroff2021,Falcaro2014, yamabayashi2018scaling,yamauchi2022quantum}. 
These advantages have long motivated interest in exploring molecular spin–qubit platforms, pursuing better coherence properties, enhanced robustness and more environmentally suitable form factors. Our work bridges these principles into practice, establishing a powerful and scalable approach for room-temperature quantum sensing.

\section{Acknowledgment}
We thank Ashok Ajoy, Shimon Kolkowitz, Max Attwood, Andrew Stasiuk, Bo Xing, Keyuan Zhong and Takuya Isogawa for helpful discussions. This work was supported in part by the National Science Foundation under Grants No. PHY2317134 (Center for Ultracold Atoms). B. L. thanks MathWorks for their support in the form of a Graduate Student Fellowship. The opinions and views expressed in this publication are from the authors and not necessarily from MathWorks.
\section{Author Contributions}
B.L., Y.Q., G.H., and P.C. conceived the idea. B.L. and G.H. implemented experiments with assistance from A.U and J.Y., and analyzed the data. Y.Q. and P.H. were responsible for pentacene-naphthalene synthesis. P.C. supervised the project. All authors discussed the results. 

\section*{Conflicts of interest}
The authors declare no conflict of interest.
\bibliographystyle{achemso}
\bibliography{pentacene_rabi_ref}

\end{document}


\begin{center}
{\bf \large{Supplemental Materials:}}\\
{\bf \Large{Robust AC vector sensing at zero-field with pentacene}}
\end{center}

\author{Boning Li}
    \thanks{These authors contributed equally.}
    \affiliation{Department of Physics, Massachusetts Institute of Technology, MA 02139, USA}
    \affiliation{Research Laboratory of Electronics, Massachusetts Institute of Technology, Cambridge, MA 02139, USA}
\author{Garrett Heller}
    \thanks{These authors contributed equally.}
    \affiliation{Research Laboratory of Electronics, Massachusetts Institute of Technology, Cambridge, MA 02139, USA}
    \affiliation{Department of Electrical Engineering and Computer Science, Massachusetts Institute of Technology, MA 02139, USA}

\author{Jungbae Yoon}
    \affiliation{Research Laboratory of Electronics, Massachusetts Institute of Technology, Cambridge, MA 02139, USA}
\author{Alexander Ungar}
    \affiliation{Research Laboratory of Electronics, Massachusetts Institute of Technology, Cambridge, MA 02139, USA}
    \affiliation{Department of Electrical Engineering and Computer Science, Massachusetts Institute of Technology, MA 02139, USA}
\author{Hao Tang}
    \affiliation{Department of material science and engineering, Massachusetts Institute of Technology, Cambridge, MA 02139, USA}
\author{Guoqing Wang}
    \affiliation{Department of Physics, Peking University, Beijing, 100871, P.R.China}
\author{Patrick Hautle}
    \affiliation{Paul Scherrer Institute, 5232 Villigen, Switzerland}
\author{Yifan Quan}
    \affiliation{Department of Chemistry, University of Pennsylvania, Philadelphia, PA 19104, USA}
\author{Paola Cappellaro}%
    \email{pcappell@mit.edu}
    \affiliation{Department of Physics, Massachusetts Institute of Technology, MA 02139, USA}
    \affiliation{Research Laboratory of Electronics, Massachusetts Institute of Technology, Cambridge, MA 02139, USA}
    \affiliation{Department of Nuclear Science and Engineering, Massachusetts Institute of Technology, Cambridge, MA 02139, USA}

\maketitle

\tableofcontents
\section{Experiment setup}
The experiments were performed on a home-built confocal microscope. A 532\,nm laser (SPROUT, Lighthouse Photonics) was sent through an acousto-optic modulator (AOM; Isomet M113-aQ80L-H) for fast switching, passed through a polarizing beam splitter, and then focused onto the sample using an oil-immersion objective (Thorlabs N100X-PFO, Nikon Plan Fluor). The fluorescence was collected through the same objective, separated from the excitation path by a dichroic mirror (Chroma NC338988), filtered by a 549\,nm long-pass filter, and detected using a single-photon counting module (PerkinElmer SPCM-AQRH-14). Microwave control pulses were generated and programmed using an arbitrary waveform generator (Tektronix AWG5014C), mixed with a sine wave from a signal generator (Stanford Research Systems SG386) using an IQ mixer, amplified (ZHL-30W-252-S+), and delivered to the sample via a home-built coplanar waveguide (CPW). Both sample and the CPW are mounted on a 3D-piezo scanner (NPoint LC400). 

\section{Pentacene-naphthalene sample preparation}
High quality pentacene-naphthalene single crystals were prepared from zone-refined naphthalene-h$_8$ doped with custom synthesized pentacene-d$_{14}$ (ISO-TEC, Sigma-Aldrich Group) using a self-seeding Bridgman growth technique~\cite{Quan2019}. The naphthalene starting material ($>99\%$) was purchased from Sigma-Aldrigh Group and further purified using zone refinement technique with more than 200 cycles. The resulting long proton NMR $T_1$ confirm the high chemical purity of the host material \cite{Steiner2023}. The pentacene concentration in the crystal used for the measurements was determined by optical spectroscopy to be approximately $4\times10^{-5}$ mol/mol (pentacene over naphthalene)~\cite{Quan2023}.
\section{Contrast calibration}
\subsection{Triplet-state lifetime and optical detection contrast characterization\label{sectionS:lifetime}}
\begin{figure}
\centering
\includegraphics[width=1\linewidth]{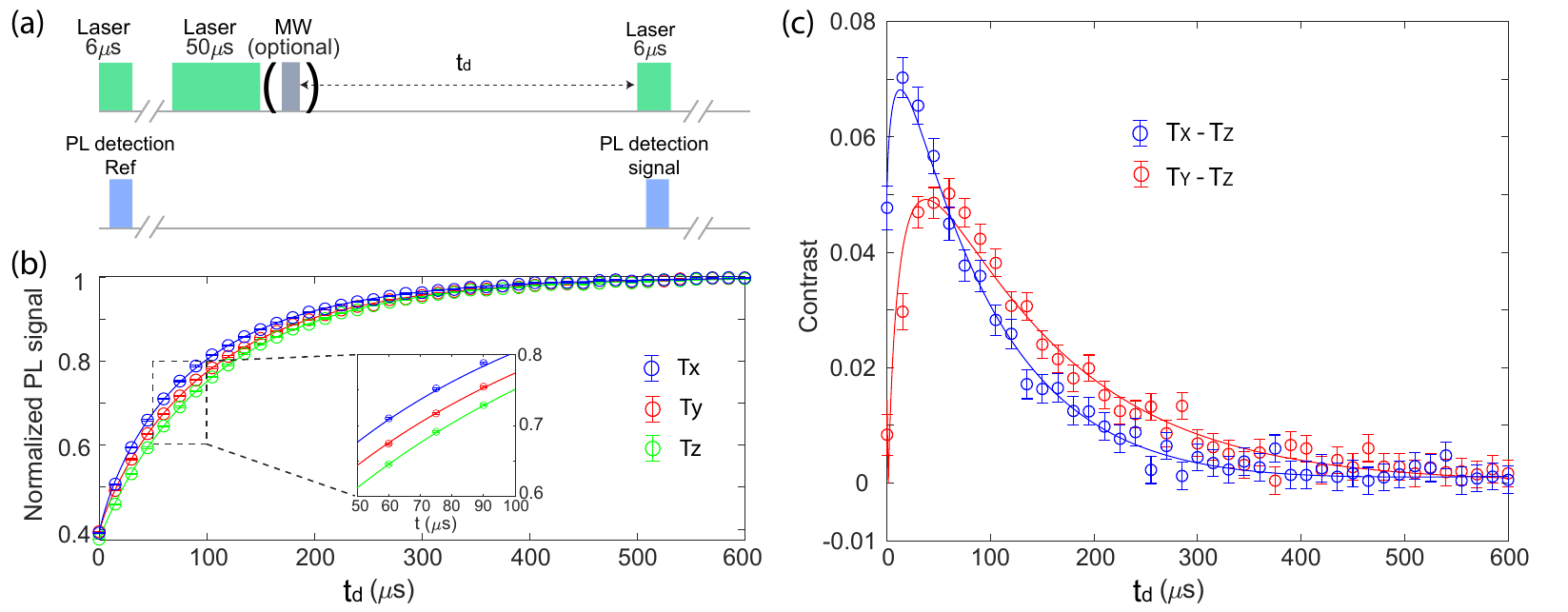}
\caption{\textbf{Population relaxation of metastable triplet states and resulting optical contrast.} 
(a) Pulse sequence used to measure population relaxation. A sufficiently long wait time ($\sim500~\mu$s) is inserted between the reference and signal measurements, and between experimental repetitions, to fully reset the system into $S_0$. A short microwave pulse is applied when initializing the triplet manifold into $T_Z$ or $T_Y$ prior to the dark interval. 
(b) Measured population relaxation starting from initial preparation in $T_X,$ $T_Y,$ or $T_Z$. The normalized photoluminescence (PL) signals are fitted with stretched-exponential decays, yielding effective decay times of $86$, $109$, and $102~\mu\mathrm{s}$ for $T_X$, $T_Y$, and $T_Z$, respectively. These differences in decay rates enable spin-state discrimination. The inset highlights that a dark interval of $50$–$100~\mu$s yields several-percent contrast between the PL signals. 
(c) Optical readout contrast for using either the $T_X\!\leftrightarrow\! T_Z$ or $T_Y\!\leftrightarrow\! T_Z$ transition as a qubit, obtained from Eq.~\eqref{eqS:contrast}. For both cases, the contrast peaks at $T_d \approx 30~\mu$s for $T_X\leftrightarrow T_Z$ transition and $T_d \approx 47~\mu$s for $T_Y\leftrightarrow T_Z$ transition under our experimental conditions.}

\label{figS:lifetime}
\end{figure}

In the main text, we emphasized that both the optical readout contrast and the time required for population reset are determined by the lifetimes of the triplet sublevels. Figure~S\ref{figS:lifetime}(a,b) presents the measured triplet lifetimes, and Figure~S\ref{figS:lifetime}(a) also shows the pulse sequence. The photoluminescence (PL) signal is proportional to the population in the ground state $S_0$. At the start of the experiment, a short laser pulse ($\sim6~\mu$s) is used to establish a reference PL level corresponding to all population residing in $S_0$.

As discussed in the main text, spin-dependent intersystem crossing (ISC) predominantly transfers population into the $T_X$ sublevel following $\sim50~\mu$s optical excitation, theoretically reaching the population of $\sim91\%$~\cite{van1980epr}. To determine its lifetime, we monitor the recovery of the $S_0$ population after a dark interval $t_d$. As $T_X$ relaxes back to the ground state, the PL signal correspondingly increases. All signals are normalized to the reference PL level corresponding to complete ground-state population. The lifetimes of the $T_Z$ and $T_Y$ sublevels were measured by first applying fast ($\sim 100~\mathrm{ns}$) microwave pulses immediately after optical excitation to transfer the population from $T_X$ into the desired sublevel. For the $T_Z$ lifetime, a single $T_X \rightarrow T_Z$ pulse was applied; for the $T_Y$ lifetime, a pulse sequence $T_X \rightarrow T_Z \rightarrow T_Y$ was used. Following this state preparation, the system relaxes to the ground state in the dark for a variable delay time $t_d$, after which the photoluminescence was measured to monitor the relaxation. 

During the dark time $t_d$, the population dynamics are described by the rate equation
\begin{equation}
\frac{d}{dt}
\!\left[
    \begin{matrix}
        \rho_X\\[2pt]
        \rho_Y\\[2pt]
        \rho_Z\\[2pt]
        \rho_{S_0}
    \end{matrix}
\right]
=
\left[
    \begin{matrix}
        -(k_{X}+w_{XY}+w_{XZ}) & w_{YX} & w_{ZX} & 0 \\
        w_{XY} & -(k_{Y}+w_{YX}+w_{YZ}) & w_{ZY} & 0 \\
        w_{XZ} & w_{YZ} & -(k_{Z}+w_{ZY}+w_{ZX}) & 0 \\
        k_X & k_Y & k_Z & 0
    \end{matrix}
\right]
\!\left[
    \begin{matrix}
        \rho_X\\[2pt]
        \rho_Y\\[2pt]
        \rho_Z\\[2pt]
        \rho_{S_0}
    \end{matrix}
\right],
\label{eqS:rate_equation}
\end{equation}
where $k_\mu$ ($\mu = X,Y,Z$) denotes the decay rate from triplet sublevel $T_\mu$ to the ground state, and $w_{\mu\nu}$ describes the transition rate between triplet sublevels arising from spin-lattice relaxation processes. The full set of rates for pentacene in a naphthalene host at room temperature has not yet been established~\cite{Mena2024,takeda2002zero}. In our experiments, when the system is initialized predominantly in a single triplet sublevel, the recovery of the ground-state population $S_0$ (and thus the normalized PL signal) can be well approximated by a stretched exponential,
\begin{equation}\label{eq:rdecay}
    PL_{\mu}(t_d)
    = 1 - (1-r)\, e^{[-\Gamma_\mu t_d]^{j_\mu}},
\end{equation}
where $r$ ($0<r<1$) accounts for the residual ground-state population immediately after optical pumping. Under laser powers that avoid damage to the pentacene–naphthalene crystal, we obtain $r\approx 0.4,$ so $1-r=0.6$ represents the optical pumping efficiency. This is significantly higher than that in typical transient triplet EPR experiments, where bulk excitation is required, because confocal imaging enables much higher local laser intensities~\cite{Quan2023}.  Here, $\Gamma_\mu$ is the effective decay rate of sublevel $T_\mu$, and the stretch exponent extracted from fitting, $j_\mu \approx 0.8$, captures the multichannel relaxation pathways. Consistent with the known hierarchy of decay rates, the results show that $\Gamma_X>\Gamma_Y>\Gamma_Z$.

The PL difference between any two triplet sublevels $\mu$ and $\nu$ determines the optical readout contrast. For an effective qubit encoded in either $T_X\leftrightarrow T_Z$ or $T_Y\leftrightarrow T_Z$, the corresponding contrast is
\begin{equation}
C(t_d)
=
\left|
\frac{PL_\mu - PL_\nu}{PL_\mu}
\right|
=
\frac{(1-r)\!\left[e^{[-\Gamma_\mu t_d]^{j_\mu}} - e^{[-\Gamma_\nu t_d]^{j_\nu}}\right]}
     {1-(1-r)e^{[-\Gamma_\mu t_d]^{j_\mu}}},
\label{eqS:contrast}
\end{equation}
which is plotted in Fig.~S\ref{figS:lifetime} (c). 

\subsection{Excitation light polarization dependence}

\begin{figure}
    \centering
    \includegraphics[width=0.5\linewidth]{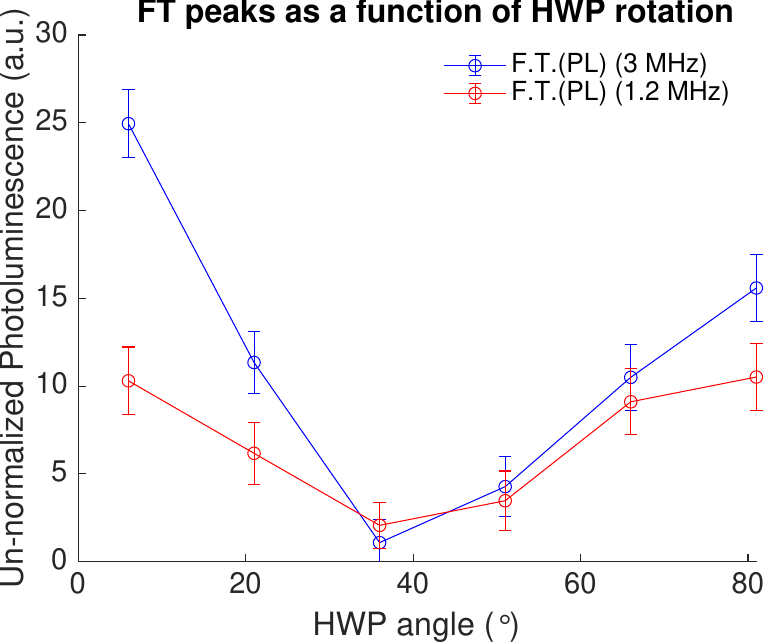}
\caption{\textbf{Polarization dependence of photoluminescence amplitude.} 
The excitation laser is linearly polarized and rotated using a half-wave plate (HWP) before the objective. The horizontal axis shows the rotation angle of the half-wave plate, which determines the incident polarization. The vertical axis shows the unnormalized Rabi-oscillation amplitudes of the $T_X \leftrightarrow T_Z$ transition; the two traces (red and blue) correspond to the two crystallographic orientations of pentacene. Unlike other measurements in this work, the photoluminescence is not normalized, allowing direct visualization of the total fluorescence variation, which reflects the number of triplets excited under each polarization condition.}

    \label{figS:pol_tune}
\end{figure}

Absorption of pentacene is polarization dependent~\cite{Quan2021}. Because the two crystallographic orientations of pentacene are inequivalent, their optical absorption also differs for a given polarization, which in turn affects the initial population pumped into the triplet manifold. Experimentally, we illuminate the sample with linearly polarized light and observe that the Rabi oscillation amplitudes of the two orientations (corresponding to the two distinct Rabi frequencies of the $T_X\!\leftrightarrow\!T_Z$ transition) vary systematically as the excitation polarization is rotated, as shown in Fig.~S\ref{figS:pol_tune}. Since both orientations exhibit similar polarization dependence, we choose the excitation polarization that maximizes the total photoluminescence, thereby maximizing the effective number of initialized spins in the detection volume.

\section{Details of sensing performance analysis}
\subsection{Spectrum broadening effects}
\begin{figure}
\centering
\includegraphics[width=1\linewidth]{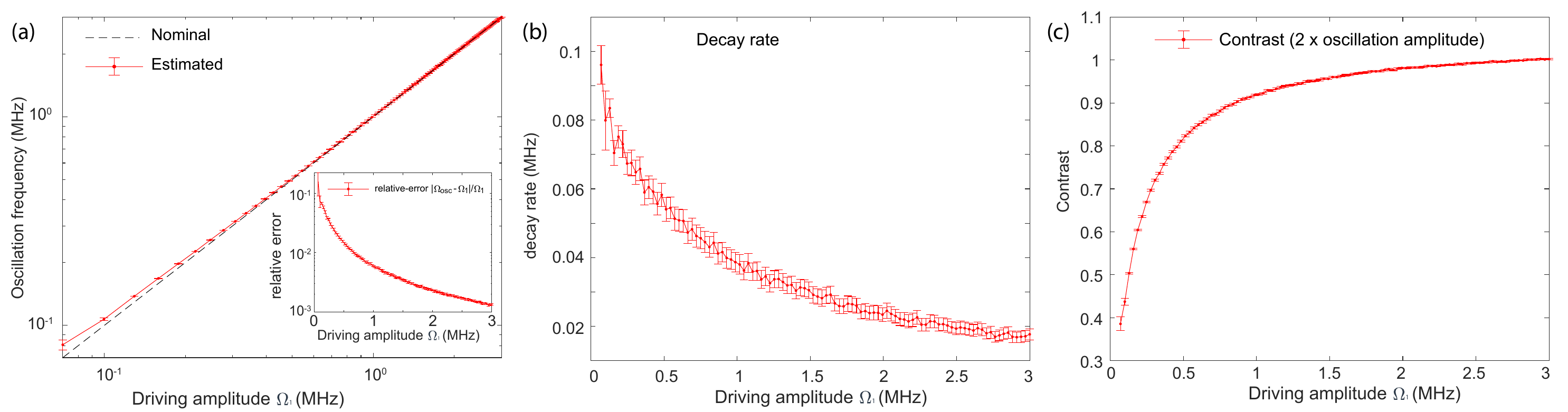}
\caption{\textbf{Numerical simulation of hyperfine-induced spectral broadening effects.} 
(a) Fitted Rabi oscillation frequency from numerical simulations assuming a Lorentzian detuning distribution with FWHM equal to the measured ODMR linewidth ($0.6~(2\pi)\mathrm{MHz}$), shown together with the nominal driving amplitude $\Omega_1$. In the weak-driving regime, the fitted oscillation frequency deviates from the nominal value. Inset: relative error between the fitted Rabi oscillation frequency and the nominal drive amplitude $\Omega_1$. 
(b) Numerically evaluated decay rate of the Rabi oscillation under the same detuning assumptions.
(c) Numerically evaluated contrast of the Rabi oscillation as twice the fitted oscillation amplitude. The increased decay rate and decreased contrast indicate sensitivity degrades at low frequencies.}

\label{figS:boradening_effect}
\end{figure}
As mentioned in the main text, the photoexcited triplet of deuterated pentacene in naphthalene is hyperfine coupled to 16 deuterons within the molecule, which produces an almost continuous distribution of electronic transition frequencies, as revealed by the measured ODMR linewidth. In the rotating frame, where the microwave driving is $H_{\mu w} = (\Omega_{1}/2)\sigma_X$, the finite linewidth corresponds to an inhomogeneous (quasi-static) distribution of detunings $\delta\omega$ between the molecular transition frequencies and the applied microwave drive. Each detuning produces an effective Rabi frequency
\[
\Omega_R(\delta\omega)=\sqrt{\Omega_1^2+\delta\omega^2}.
\]
The ensemble-averaged signal is therefore a superposition of Rabi oscillations with different $\Omega_R$:
\begin{equation}
    P(t)
    = 1-\int \mathrm{d}\Omega_R\, f(\Omega_R)\,
    \frac{\Omega_1^2 \sin^2(\Omega_R t/2)}{\Omega_R^2},
    \label{eq:P_ensemble}
\end{equation}
where $f(\Omega_R)$ denotes the distribution of effective Rabi frequencies generated from the distribution of $\delta\omega$.  From Eq.~\eqref{eq:P_ensemble}, three consequences directly could impact sensing performance.  First, because $\Omega_R > \Omega_1$ for any nonzero detuning, using the measured Rabi oscillation frequency $\Omega_{\mathrm{osc}}$ to infer the field amplitude leads to a systematic overestimation. Second, as the detuning dominates the Rabi oscillation, the contrast reduced due to the small Rabi oscillation amplitude $\Omega_1^2/\Omega_R^2$. Thirdly, the finite spread of $\Omega_{\mathrm{osc}}$ produces inhomogeneous decay of the ensemble-averaged Rabi signal. 

We numerically investigated this effect by simulating the Rabi oscillation time trace based on Eq.~\ref{eq:P_ensemble}, over an ensemble average of the detuning distribution by randomly sampling it from a Lorentzian distribution (zero centered, FWHM given by the ODMR linewidth $0.6~(2\pi)\mathrm{MHz}$). Then we fit the time trace with a damped oscillation function to extract the oscillation frequency $\Omega_{\mathrm{osc}}$ as an estimation of the driving field amplitude $\Omega_1$, the contrast as the oscillation amplitude, and the decay rate, shown in Fig.~S\ref{figS:boradening_effect}.

\subsection{Relaxation mechanism-Rabi}
In the strong-driving limit, the splitting of the dressed spin states is predominantly determined by the drive amplitude $\Omega_1$, and decoherence is dominated by stochastic magnetic fields and fluctuations in this drive. The Hamiltonian in the lab frame is presented in Eq.~6 in the main manuscript. In the rotating frame, it becomes:
\begin{equation}
H(t)
= \frac{\Omega_{1}\!\left[1+\xi_{\Omega_1}(t)\right]}{2}\sigma_X
+ \xi_Z(t)\sigma_Z
+ \xi_{\perp}(t)\!\left[\sigma_X\cos(\omega_0 t) + \sigma_Y\sin(\omega_0 t)\right],
\label{eq:H_noise_disc}
\end{equation}
where stationary noise functions $\xi_Z(t)$ and $\xi_{\perp}(t)$ represent longitudinal and transverse magnetic-field fluctuations, respectively (dominated by the surrounding nuclear-spin bath), and $\xi_{\Omega_1}(t)$ describes relative fluctuations in the drive amplitude. The oscillatory factors $\cos(\omega_0 t)$ and $\sin(\omega_0 t)$ arise from transforming the transverse noise into the rotating frame at frequency $\omega_0$.

Each stochastic term is characterized by its autocorrelation function and corresponding power spectral density (PSD), $S_j(\nu)
= \int_{-\infty}^{\infty} \mathrm{d}t\, e^{i \nu t} \,\langle\xi_j(t)\xi_j(0)\rangle,
\  j \in \{z,\perp,\Omega_1\}$.
Quasi-static spatial inhomogeneity of the magnetic and microwave fields enter as an additional zero-frequency contribution in the corresponding spectra. Within the generalized Bloch-equation (GBE) framework~\cite{geva1995relaxation,Wang2020}, the relaxation rates of the driven spin in the rotating frame are
\begin{equation}
    \frac{1}{T_{1\rho}} = S_Z(\Omega_1) + \frac{1}{2}S_X(\omega_0),\qquad
    \frac{1}{T_{2\rho}} = \frac{1}{2T_{1\rho}} + \Gamma_{\mathrm{ph}},
    \label{eq:T1rho_T2rho_disc}
\end{equation}
where $T_{1\rho}$ describes relaxation of the dressed-state population. For slow magnetic-field fluctuations, the longitudinal noise $\xi_Z(t)$ that flips the dressed state produces a PSD $S_Z(\nu)$ centered at zero frequency, and its value at $\nu = \Omega_1$ sets $T_{1\rho}$.

The term $\Gamma_{\mathrm{ph}}$ accounts for additional phase diffusion arising from fluctuations of the field along the drive axis ($X$). To evaluate it, we consider the stochastic phase accumulated during an evolution time $T$,
\begin{equation}
\phi(T)
= \int_0^{T}\! \mathrm{d}t\,
\left[\frac{\Omega_1\,\xi_{\Omega_1}(t)}{2}
+ \xi_{\perp}(t)\cos(\omega_0 t)\right].
\end{equation}
Assuming zero-mean Gaussian noise, the decay of the coherent Rabi oscillation is determined by the phase variance via $\exp[-\langle\phi^2(T)\rangle/2]$, where
\begin{equation}
    \langle \phi^2(T)\rangle
    = \int_0^{T}\!\mathrm{d}t_1\!\!\int_0^{T}\!\mathrm{d}t_2
    \left[\frac{1}{4}
    \Omega_1^{2}\langle\xi_{\Omega_1}(t_1)\xi_{\Omega_1}(t_2)\rangle
    + \langle\xi_{\perp}(t_1)\xi_{\perp}(t_2)\rangle
      \cos(\omega_0 t_1)\cos(\omega_0 t_2)
    \right].
    \label{eq:phi2_time_disc}
\end{equation}
Using the PSD definitions, this expression becomes
\begin{equation}
    \langle \phi^2(T)\rangle
    = T^{2}\!
    \int\!\mathrm{d}\nu\,
    \left[\frac{1}{4}
    S_{\Omega_1}(\nu)
    + \frac{1}{2}S_{\perp}(\nu-\omega_0)
    \right]
    \mathrm{sinc}^{2}\!\left(\frac{\nu T}{2}\right).
\end{equation}

For amplitude noise, when $\xi_{\Omega_1}(t)$ varies slowly in time or originates from spatial inhomogeneity of the microwave field, its PSD is dominated by a zero-frequency component~\cite{Wang2020},
\[
S_{\Omega_1}(\nu)
= 2\pi\,\Omega_1^{2}\,\mathrm{Var}[\xi_{\Omega_1}]\,\delta(\nu),
\]
so the associated dephasing follows a Gaussian decay with
\[
\langle\phi^{2}(T)\rangle_{\Omega_1} \propto (\Omega_1 T)^{2}.
\]

For transverse magnetic-field variation, $\xi_{\perp}(t)$ is typically well approximated as a classical noise. In this limit the $\mathrm{sinc}^{2}$ window approaches a delta function, picking out the noise at the dressed-state transition frequency:
\[
\langle\phi^{2}(T)\rangle_{\perp}
\propto
T\, S_{\perp}(\omega_0).
\]
Thus, Gaussian amplitude noise contributes a Gaussian (quadratic-in-time) dephasing term, whereas Gaussian transverse magnetic noise contributes a Lorentzian (linear-in-time) dephasing term.

\subsection{Relaxation mechanism-Rotary-Echo}
The improved rotary--Echo (RE) protocol addresses these limitations by introducing a strong control drive $\Omega_{\mathrm{c}}$ with periodic $\pi$ phase inversions, applied simultaneously with the weaker target field to be sensed ($\Omega_{\mathrm{t}}$). Under this control scheme, the microwave part in the rotating frame is
\begin{equation}
H_{\mu w}
= \frac{1}{2}\!\left[
\Omega_{\mathrm{t}}\big(1+\xi_{\Omega_t}(t)\big)
+ \Omega_{\mathrm{c}}\, s(t)\big(1+\xi_{\Omega_c}(t)\big)
\right]\sigma_X,
\label{eq:H_RCPMG_disc}
\end{equation}
where $\xi_{\Omega_t}(t)$ and $\xi_{\Omega_c}(t)$ denote relative fluctuations of the target and control amplitudes, respectively, and $s(t)=\pm 1$ implements the periodic inversions with period $2\tau$. 

This scenario is slightly different from a well-defined Floquet configuration, in which no continuous drive is present. To address this, one can transform the system into a toggling rotating frame defined by $U = \exp\left[-i\,\frac{\Omega_t}{2}\sigma_X\right]$.
In this frame, the Hamiltonian including driving and noise terms becomes:
\begin{equation}
\begin{aligned}
      H' =& \frac{1}{2}\!\left[
\Omega_{\mathrm{t}}\xi_{\Omega_t}(t)
+ \Omega_{\mathrm{c}}\, s(t)\big(1+\xi_{\Omega_c}(t)\big)
\right]\sigma_X,\\
&+ \xi_Z(t)[\sigma_Z\cos(\Omega_tt)+\sigma_Y\sin(\Omega_tt)]\\
\end{aligned}
\end{equation}
Here we employ the rotating-wave approximation. The transverse noise term in Eq.~\eqref{eq:H_noise_disc} is neglected here for its marginal contribution to the relaxation rate set by $S_{\perp}(\omega_0)$. Here for ease of following discussion, one can define the new fluctuation terms $\xi_{Z1}(t) = \xi_{Z}\cos(\Omega_tt)$ and $\xi_{Z2}(t) = \xi_{Z}\sin(\Omega_tt)$, and both drives flips of the dressed state whose splitting is set by $\Omega_cs(t)$ along $X$ direction. The power spectral density for $\xi_{Z1}$ and $\xi_{Z2}$ can be easily calculated as 
\begin{equation}
    S_{Zi}(\nu) = \frac{1}{4}\left[S_Z(\nu-\Omega_t)+S_Z(\nu+\Omega_t)\right],
\end{equation}
which is identical for $i = 1,2$ and shifted by the rotating frame frequency. 

It is now clear that the system forms a well-defined Floquet configuration: the periodic drive $s(t)\Omega_c$ generates sidebands at resonance frequencies $
\nu = \Omega_c \pm \frac{(2m+1)\pi}{\tau}, \  m = 1,2,\ldots$,
where each sideband carries a weight proportional to $\frac{1}{(2m+1)^2}$.
 ~\cite{Aiello2013, hirose2012continuous,ivanov2021floquet}. Therefore, the longitudinal relaxation given by the spectrum component of the noise $S_{Z1}(\nu)$ and $S_{Z2}(\nu)$ on resonance with the sideband transitions:
\begin{widetext}
\begin{equation}
\begin{aligned}
    \left.\frac{1}{T_{1\rho}}\right|_{\mathrm{R\!-\!SE}}
    &= \sum_{m=-\infty}^{\infty}
    \frac{1}{2(2m+1)^2}
    \Big[
    S_{z}\!\left(\Omega_{\mathrm{t}}+\Omega_{\mathrm{c}}
    -\frac{(2m+1)\pi}{\tau}\right)
    +
    S_{z}\!\left(-\Omega_{\mathrm{t}}+\Omega_{\mathrm{c}}
    -\frac{(2m+1)\pi}{\tau}\right)
    \Big] \\
    &\approx \sum_{m=-\infty}^{\infty}
    \frac{1}{(2m+1)^2}\,
    S_{z}\!\left(\Omega_{\mathrm{c}}(1-\frac{(2m+1)\pi}{\Omega_{\mathrm{c}}\tau})\right),
\end{aligned}
\label{eq:T1rho_RCPMG_Floquet}
\end{equation}
\end{widetext}
where the approximation holds in the strong-control–field limit, $\Omega_{\mathrm{c}} \gg \Omega_{\mathrm{t}}$. 

The additional dephasing $\Gamma_{ph}|_{RE}$ induced by the fluctuation of the control channel is
\begin{equation}
    \phi_{\mathrm{c}}(T) 
    = \int_0^T \mathrm{d}t\, \Omega_{\mathrm{c}}\,\xi_{\Omega_c}(t)\,s(t),
\end{equation}
and the corresponding phase variance is
\begin{equation}
    \langle \phi_{\mathrm{c}}^2(T)\rangle 
    = \int_0^T \mathrm{d}t_1\!\int_0^T \mathrm{d}t_2\,
    \Omega_{\mathrm{c}}^2\langle\xi_{\Omega_c}(t_1)\xi_{\Omega_c}(t_2)\rangle 
    s(t_1)s(t_2) 
    =\int\mathrm{d}\nu S_{\Omega_{\mathrm{c}}}(\nu)F(\nu),
\end{equation}
\begin{equation}
    F(\nu) = \frac{4}{\nu^2}\tan^2(\frac{\nu\tau}{2})\sin^2(N\nu\tau).
\end{equation}
where $2N$ is the total pulse number ($T = 2N\tau$). In the frequency domain, the modulation $s(t)$ gives rise to a CP-like filter function ($F(\nu)$) with peaks at odd harmonics $\nu\tau = (2m+1)\pi$~\cite{alvarez2011measuring,bylander2011noise}. As the static spatial inhomogeneity or slow temporal fluctuation in the control amplitude induces the part of $S_{\Omega_c}(\nu)$ that is strongly peaked near zero frequency. This inhomogeneity dominates $S_{\Omega_1}(0)$ in the simple Rabi case; however, its contribution at the finite harmonics is strongly suppressed. 

\section{Determining the confocal volume of our set-up}
In the main text, we highlighted the importance of the high spin density achievable in molecular systems. On one hand, a larger number of sensing spins improves the magnetic sensitivity, which scales as $\eta \propto 1/\sqrt{N_s}$, where $N_s = \rho_s V$ is the number of active spins determined by the spin density $\rho_s$ and the detection volume $V$. On the other hand, achieving high spatial resolution, or avoiding coherence degrading over inhomogeneity requires a small detection volume, which reduces $N_s$ and therefore degrades sensitivity. To compare different sensing platforms on equal footing, independent of the specific measurement volume, it is convenient to use a volume-normalized sensitivity,
\begin{equation}
    \eta_V \equiv \eta \sqrt{V},
\end{equation}
which captures the intrinsic performance per unit detection volume and directly reflects the role of spin density.

\begin{figure}
    \centering
    \includegraphics[width=1\textwidth]{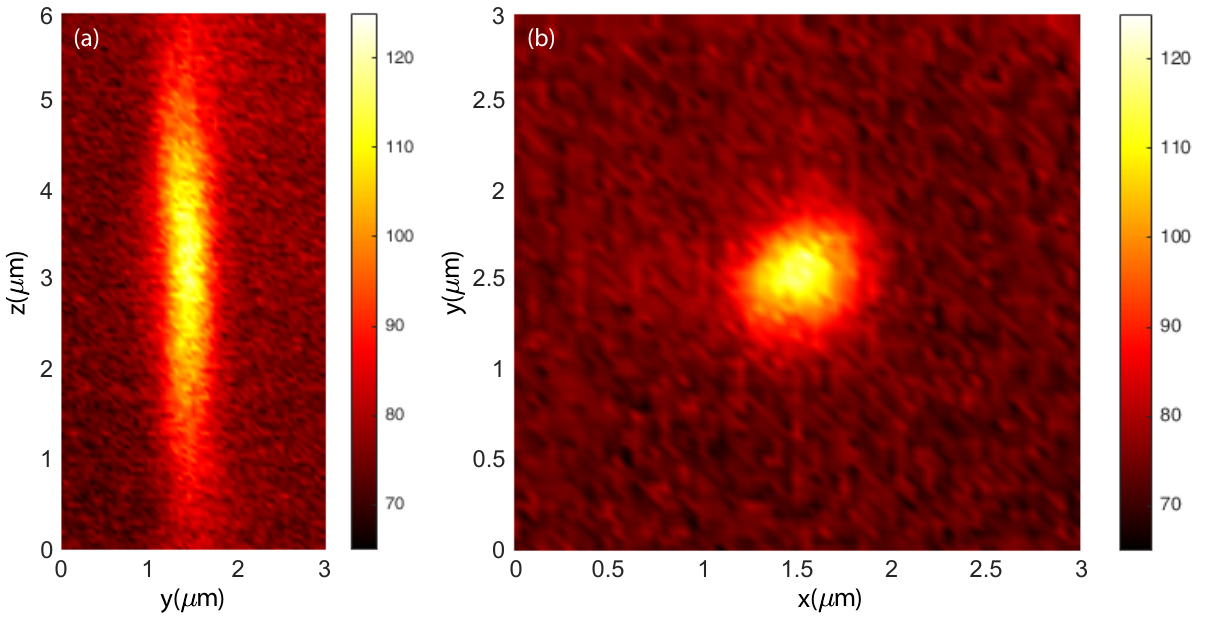}
 \caption{\textbf{Confocal images of a single NV center.} 
Using the same confocal microscope employed for all measurements in this work, we record the photoluminescence (PL) from an isolated NV center (color scale: kilocounts/s). The beam propagates along the $z$-axis.
(a) Axial ($z$) scan showing the depth of focus. 
(b) Lateral ($x$–$y$) scan in the imaging plane. 
These measurements are used to calibrate the diffraction-limited optical detection volume of the setup.}
    \label{figS:confocalNV}
\end{figure}

We conduct our experiments using a confocal microscope, in which the optical detection volume is set by the diffraction-limited point-spread function (PSF) of the objective. To determine this volume experimentally, we perform a three-dimensional scan of the PL from a single NV center, as shown in Fig.~S\ref{figS:confocalNV}. Assuming a Gaussian PSF, we extract the characteristic widths along each axis by measuring the distance from the maximum PL intensity to the $1/e$ level. The lateral (in-plane) radii are approximately $0.35~\mu\mathrm{m}$, consistent with the diffraction limit for our objective, while the axial extent (depth of focus) is about $2~\mu\mathrm{m}$. These dimensions correspond to an effective detection volume of 
\[
V \approx \frac{4}{3}\pi (0.35~\mu\mathrm{m})^2 (2~\mu\mathrm{m}) \sim 1~\mu\mathrm{m}^3.
\]
Given the molecular density of pentacene in the naphthalene host, this volume contains approximately $10^{5}$ pentacene molecules contributing to the signal. 

Therefore, by normalizing the measured sensitivity to the optical detection volume, we obtain a volume-normalized sensitivity of  $\eta_V \sim 1~\mu\mathrm{T}\,\mu\mathrm{m}^{3/2}/\sqrt{\mathrm{Hz}}$ for the pentacene–naphthalene sensor. 

The sensitivity per unit volume $\eta_V$ for pentacene can be further improved, as we did not fully leverage the $\sim 10^{5}$ spins within the detection volume. Our optical pumping ratio $r$ (Section~\ref{sectionS:lifetime}) is only $0.6$, meaning that merely $60\%$ of the available spins contribute to the signal. This limitation primarily arises from photoheating constraints that prevented us from using higher optical intensities. A straightforward route to improvement is to increase the pumping efficiency. For example, Ref.~\cite{Quan2023} reports that using a $552~\mathrm{nm}$ laser rather than the standard $532~\mathrm{nm}$ excitation provides a $\sim 1.25\times$ enhancement in pentacene absorption. This increased absorption would allow the use of lower laser power or shorter pulses while still achieving higher initialization fidelity, thereby improving $\eta_V$.

We note that under our experimental conditions, where approximately $60\%$ of the pentacene molecules are photoexcited, the dipolar interactions between electron spins remain weak ($\sim 10~\mathrm{kHz}$) given their average separation of $\sim 20~\mathrm{nm}$, and therefore have only a marginal influence on the spin dynamics and coherence. At higher electron-spin densities, however, such interactions can become a relevant source of decoherence~\cite{li2025exploring,li2025quantum}.

\section{Example data for the $T_Y\leftrightarrow T_Z$ transition}
\begin{figure}[htbp]
    \centering
    \includegraphics[width=1\textwidth]{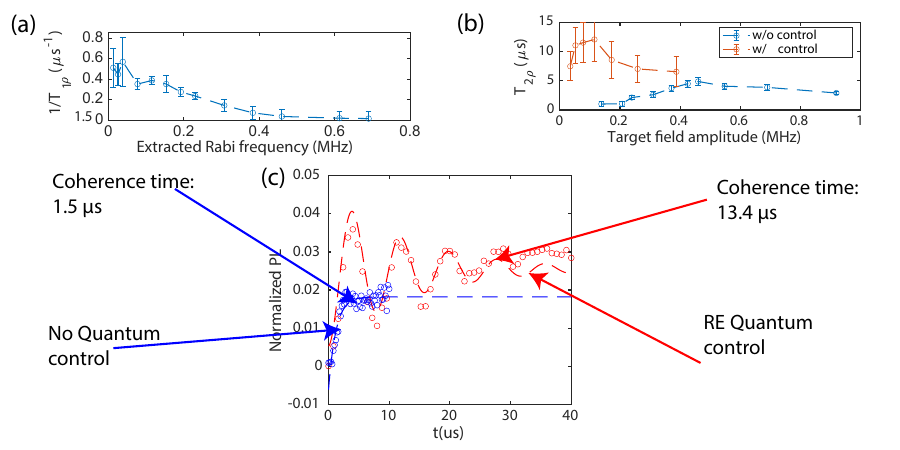}
\caption{\textbf{$T_{1\rho}$ spectrum and coherence of the $T_Y \leftrightarrow T_Z$ transition.} 
(a) Longitudinal relaxation rate $T_{1\rho}^{-1}$ measured using the same pulse sequence as in the main text, yielding the longitudinal magnetic fluctuation power spectral density $S_Z(\Omega)$. 
(b) Comparison of the driven-state coherence time with and without improved RE control as a function of the target-field amplitude. Similar to the $T_X \leftrightarrow T_Z$ transition, the RE sequence enhances the coherence time by nearly an order of magnitude in the weak-field regime. 
(c) Example Rabi oscillations with and without RE control, illustrating the extended coherence achieved under RE.}
    \label{figS:t1t2yz}
\end{figure}

In the main text, we presented the coherence of the Rabi-sensing scheme and the enhancement achieved using the rotary–echo protocol for the $T_X \leftrightarrow T_Z$ transition. For completeness, Fig.~S\ref{figS:t1t2yz} shows analogous measurements for the $T_Y \leftrightarrow T_Z$ transition, which probes magnetic-field components along the other sensing axes. The same control-pulse parameters ($\Omega_c = 3.37 ~(2\pi) \mathrm{MHz}$,  $\tau = 300 ~\textrm{ns}$) used for the $T_X \leftrightarrow T_Z$ measurements are applied here as well. 

\section{Sensing vector AC field of a home-built coplanar waveguide}
\subsection{Electronic characterization for measurements at different resonance frequencies}
In our experiments, the electronic spin transitions are driven at two distinct microwave frequencies. Because the microwave electronics and the coplanar waveguide (CPW) exhibit frequency-dependent transmission, the delivered magnetic-field amplitude differs between these frequencies. To accurately reconstruct the AC vector field, a normalization factor accounting for the frequency response of the electronics is therefore required.

Microwave excitation is delivered through a home-built CPW. Its frequency response was characterized using a vector network analyzer (VNA) via the $S_{21}$ transmission coefficient. The microwave power reaching the sample is
\begin{equation}
    P_{\mathrm{final}}(\nu)
    = P_{\mathrm{in}}(\nu)\,\left(1-\big|S_{21}(\nu)\big|^{2}\right),
\end{equation}
where $P_{\mathrm{in}}(\nu)$ is the power entering the CPW at frequency $\nu$. The input power $P_{\mathrm{in}}(\nu)$ itself also has a frequency dependence originating from upstream microwave components such as the preamplifier, microwave switch, and power combiner, which were independently characterized using a spectrum analyzer.

Combining these characterizations on our setup, the microwave power delivered to the sample at the two relevant resonance frequencies satisfies
\begin{equation}
    P_{\mathrm{final}}(1437~(2\pi)\mathrm{MHz})
    = (1.13)^{2} \,
      P_{\mathrm{final}}(1341~(2\pi)\mathrm{MHz}).
\end{equation}
This calibration factor allows us to properly renormalize the magnetic-field amplitudes extracted from Rabi measurements at the $\ket{T_Y}\!\leftrightarrow\!\ket{T_Z}$ transition ($1341 ~(2\pi)\mathrm{MHz}$) to those at the
$\ket{T_X}\!\leftrightarrow\!\ket{T_Z}$ transition ($1437 ~(2\pi)\mathrm{MHz}$), ensuring consistent reconstruction of the AC field vector.
\begin{figure}
    \centering
    \includegraphics[width=1\linewidth]{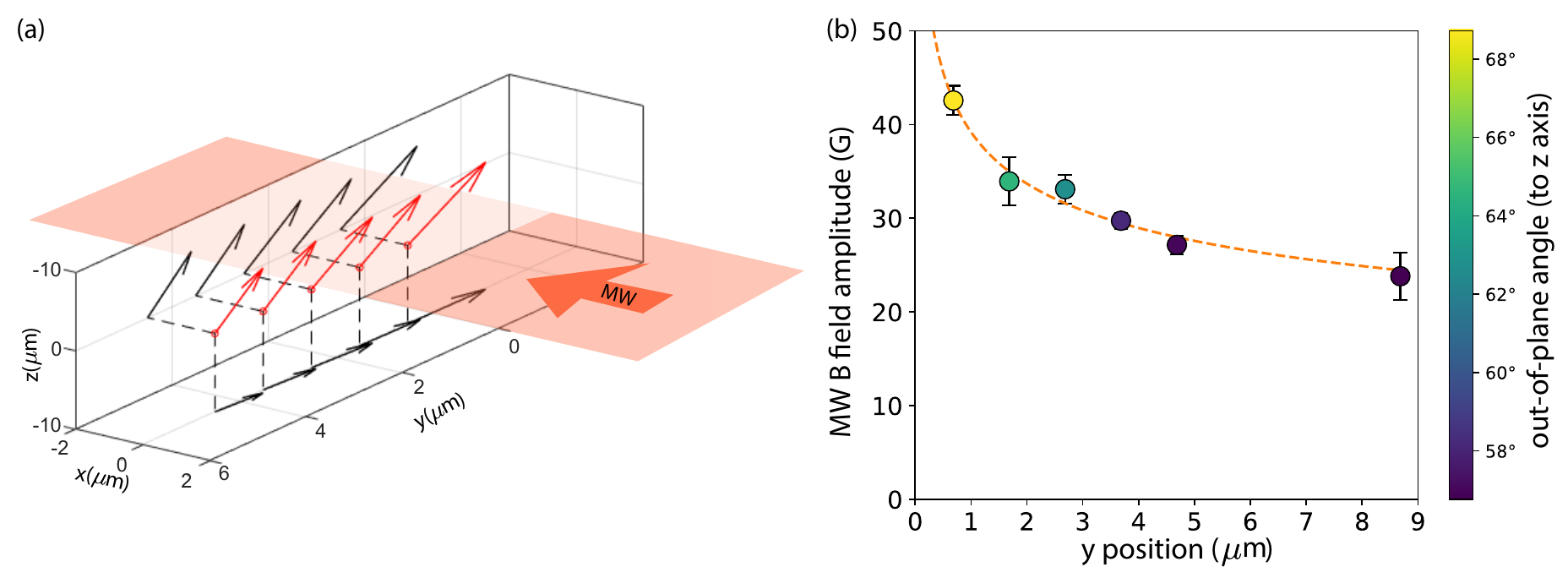}
\caption{\textbf{Vector AC field reconstruction of the CPW stripline.} 
Zero-field vector magnetometry based on Rabi oscillations is applied to map the microwave field generated by the coplanar-waveguide (CPW) stripline.  
(a) Geometry of the CPW and reconstructed local AC magnetic-field vectors at measurement points located $\sim 10~\mu$m above the stripline surface, relative positions are consistent with those in the maintext.  
The black lines show the projections of the reconstructed vectors onto the $zx$ and $xy$ planes, where the coordinates are defined by the CPW.  
(b) Reconstructed field amplitude and out-of-plane angle (relative to the $z$-axis) at each measurement point.}

    \label{figS:Bgrid}
\end{figure}

\subsection{Vector microwave field reconstruction}
The home-built CPW used for microwave delivery is a Ti(10~nm)/Cu(500~nm) stripline [Fig.~1(c) in the main text and Fig.~S\ref{figS:Bgrid} (a)] with a signal-line width of $75~\mu$m. Ground planes on both sides are separated from the signal line by $25~\mu$m, which also defines the optical window through which the excitation laser is focused and fluorescence is collected. The pentacene–naphthalene crystal is positioned approximately $10~\mu$m above the CPW surface and aligned such that its crystallographic $a$-axis runs parallel to the stripline, as verified by birefringence measurements.

To map the spatial vector profile of the microwave field generated by the CPW, we perform vector Rabi sensing at six positions along the direction perpendicular to the stripline (along the crystal $b$-axis). The reconstructed magnetic-field magnitude and orientation at each location are shown in Fig.~S\ref{figS:Bgrid}.

\section{Robustness against weak DC field}
In the main manuscript, we highlight the robustness of the pentacene sensor in the presence of a small static magnetic field (e.g., the Earth's magnetic field). The zero-field spin triplet Hamiltonian is 
\begin{equation}
    H_0 = DS_X^2 + E(S_Z^2 - S_Y^2) -\frac{2D}{3}I,
\end{equation}
yielding the eigenstate are identical to the eigenstate of each spin-1 operator with eigenvalue 0:
\begin{equation}
    |T_\mu\rangle=|S_{\mu}=0\rangle, \mu = X,Y,Z.
\end{equation}
This yields the first-order perturbation of  a weak DC field term $\mathbf{B}_{\mathrm{DC}}\!\cdot\!\mathbf{S}$  vanishes for both the triplet eigenstates and their transition energies vanishes:
\begin{equation}
    \langle T_\mu | S_\nu | T_\mu \rangle \equiv 0, 
    \qquad \mu,\nu \in \{X,Y,Z\}.
\end{equation}
In contrast, for conventional solid-state quantum sensors such as NV centers, the first-order perturbation from a DC magnetic field along the quantization axis (the N--V axis) is nonzero, resulting in a linear shift of the transition energies, which could complicates the sensing and control protocols.

\begin{figure}
    \centering
    \includegraphics[width=1\linewidth]{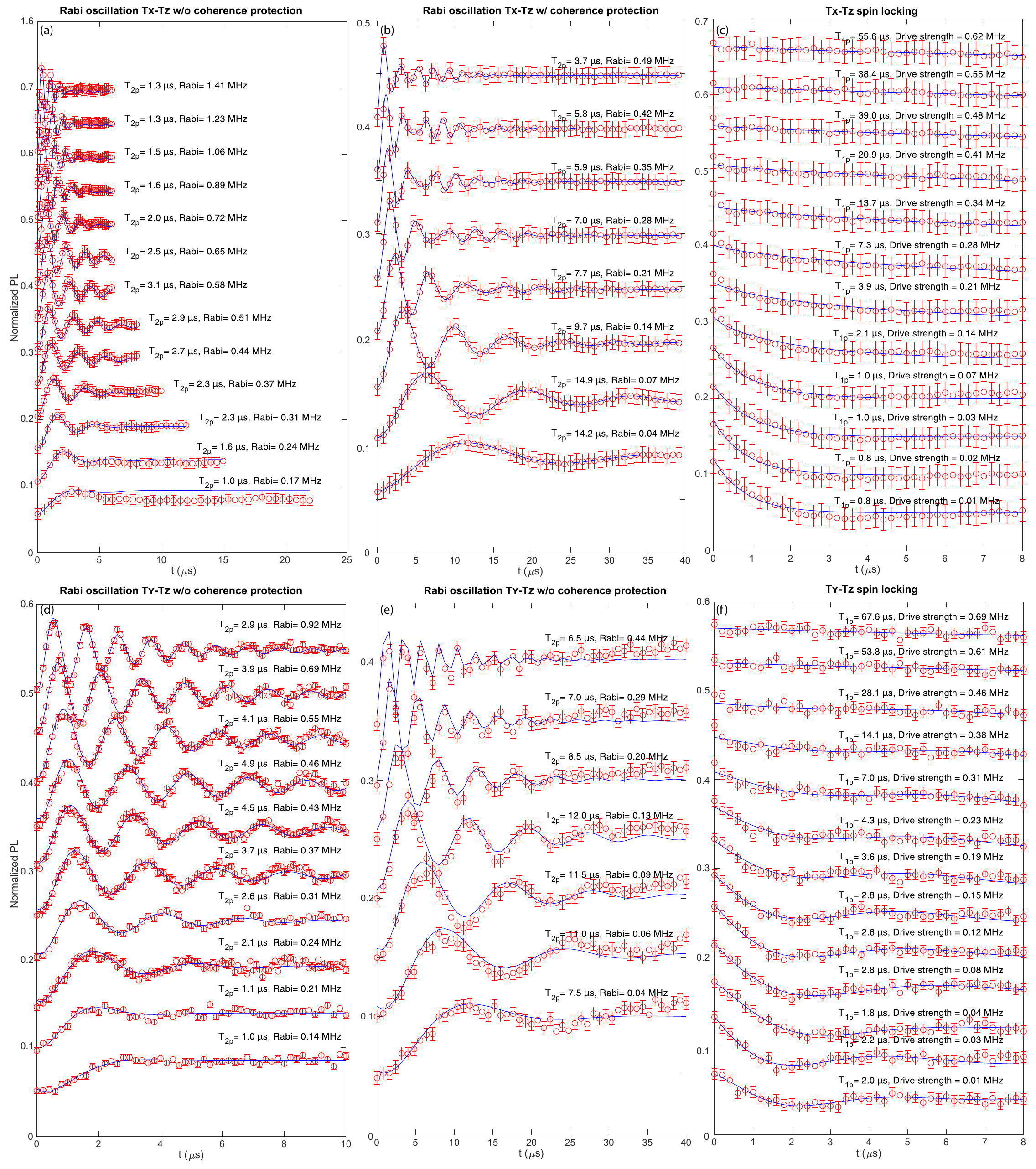}
\caption{Rabi oscillations without (a,d) and with (b,e) rotary-echo control, measured as a function of the target field amplitude, and spin-locking measurements (c,f) under varying locking pulse amplitudes. The decay rates are extracted from fits to the data (blue solid lines) and shown in the figures in the main manuscript and Supplementary Information. (a--c) correspond to the \(T_X \leftrightarrow T_Z\) transition, and (d--f) correspond to the \(T_Y \leftrightarrow T_Z\) transition. Notably, in the spin-locking measurements of the \(T_Y \leftrightarrow T_Z\) transition at low driving amplitudes, a slow oscillatory component is observed, which may arise from coherent spin transfer between the electron spin and strongly coupled nuclear spins (deuteron, hydrogen nuclear spins). Similar effects are reported in other systems~\cite{mathies2016pulsed,jain2017off,li2025exploring}. This behavior is suppressed at stronger driving amplitudes, corresponding to the operating regime of the rotary-echo sequence.
}

    \label{figS:Bgrid}
\end{figure}

\newpage\clearpage
\bibliographystyle{achemso}
\bibliography{pentacene_rabi_ref}